\documentclass[aps,prd,amsmath,amssymb,twocolumn]{revtex4}

\usepackage{graphicx}
\usepackage{dcolumn}
\usepackage{bm}
\usepackage{amsmath}
\usepackage{epstopdf}
\usepackage{amsfonts}
\usepackage{amssymb}
\usepackage{epsfig}
\usepackage{tabularx}
\usepackage{color}
\usepackage{soul}

\usepackage[colorlinks=true,citecolor=blue,urlcolor=magenta,breaklinks]{hyperref}

\newcommand{\be}{\begin{equation}}
\newcommand{\en}{\end{equation}}
\newcommand{\bea}{\begin{eqnarray}}
\newcommand{\ena}{\end{eqnarray}}

\begin{document}

\title{Accretion of matter and spectra of Binary X-ray sources in Massive Gravity}

\author{
Grigoris Panotopoulos  {${}^{a,b}$
\footnote{grigorios.panotopoulos@ufrontera.cl}
}
\'Angel Rinc\'on {${}^{c}$
\footnote{aerinconr@academicos.uta.cl}
} 
Il{\'i}dio Lopes  {${}^{a}$
\footnote{ilidio.lopes@tecnico.ulisboa.pt}
}
}

\address{
${}^a$ Centro de Astrof{\'i}sica e Gravita{\c c}{\~a}o-CENTRA, Instituto Superior T{\'e}cnico-IST, Universidade de Lisboa-UL, Av. Rovisco Pais, 1049-001 Lisboa, Portugal.   
\\
${}^b$Departamento de Ciencias F{\'i}sicas, Universidad de la Frontera, Avenida Francisco Salazar, 01145 Temuco, Chile.
\\
${}^c$ Sede Esmeralda, Universidad de Tarapac\'a, Avenida Luis Emilio Recabarren 2477, Iquique, Chile.
}


\begin{abstract}
We study low-mass binary X-ray sources involving stellar mass black holes within massive gravity. Regarding the accretion disk, we adopt the standard model for an optically thick, cool, and geometrically thin disk by Shakura-Sunyaev. For the gravitational field generated by the black hole, we consider the analogue of the Schwarzschild-de Sitter space-time of Einstein's theory in massive gravity, for which we found an additional term linear in the radial coordinate. Then, we compute the radial velocity, the energy density and the pressure as a function of the radial coordinate, and the X-ray emission's soft spectral component expected from the disk. We also investigated in detailed the impact of this new geometry. Our result indicates that by using observed spectra from confirmed X-ray binaries involving astrophysical black holes, we can put strong constraints on alternative theories of gravity.
\end{abstract}


\maketitle

\section{Introduction}

A wealth of information about physical systems is obtained observing and analyzing the light that reaches our detectors, both earth-based and space-based, from distant sources. Nowadays, high resolution, contemporary X-ray and $\gamma$-ray satellites, such as NASA's Chandra X-ray observatory and ESA's XMM Newton for instance, provide detailed spectroscopic studies. Back in 1962, Scorpius X-1 became the first X-ray source discovered outside the Solar System \cite{Giacconi:1962zz}. In the following years, X-ray binaries have proven to be important for studies of fundamental physical processes, such as mass accretion and jet formation. 

\smallskip

It is well-known that an X-ray binary is formed either when the companion star transfers matter onto the compact object by means of the inner Lagrange point or when the compact object captures mass from the wind of the companion star. Also, the transferred mass depends on i) the amount of angular momentum (that it possesses), ii) the physical processes (by which it looses angular momentum), and iii) the radiation processes (by which it cools).

\smallskip

Originally, matter accretion was considered in the context of Newtonian gravity (see \cite{hoyle1939,bondi1944,bondi1952}) and after that, it was generalized to curved space-times in \cite{michael1972}.
In particular, isothermal Bondi-like accretion has risen enthusiasm on the
galactic evolution theory community \cite{korol16a,ciotti17a,ciotti18a}. Besides, accretion process have been 
investigated in the context of General Relativity and beyond with different interests \cite{belgman1978,pretrich1988,malec1999,babichev2004,babichev2005,karkowski2006,mach2008,gao2008,
jamil 2008,giddings2008,jimenez2008,sharif2011,dokuchaev2011,bibichev2012,bhandra2012,mach2013,mach2013a,
karkowski2013,jhon2013,ganguly2014,bachivev2014,debnath2015,yang2015,jiao2017,
jiao2017a,paik2018}. 

\smallskip

Of particular interest is accretion processes in {\it{binary systems}}, where the massive star evolves faster ending as a neutron star or a black hole, whereas the companion object, being much lighter, continues to lie on the main sequence strip burning hydrogen. Matter consisting of a spherical cloud is accumulated onto the compact object coming from the companion star (the donor), and rotates in Keplerian orbits around the compact objects. As the accreting matter approaches the compact object rotates faster and faster, and it is heated up. The temperature eventually reaches millions of degrees Kelvin, and the binary star glows at the X-ray band.

\smallskip

Low-mass binary X-ray sources, involving a compact object, i.e. either a neutron star or a black hole, and a sun-like star in the main or post-main sequence phase of evolution, emit almost their entire energy at the X-ray band, and they comprise the brightest extrasolar objects. The observed spectra of such an exotic object is believed to consist of two spectral components, namely a soft and a hard one \cite{Mitsuda:1984nv}. The latter, expected from the surface of the compact object, is represented by a blackbody spectrum of $T \approx 2~keV$, whereas the former, expected from the optically thick accretion disk, is best represented by a superposition of multicolour blackbody spectra \cite{Mitsuda:1984nv}. 

\smallskip

In binary X-ray sources involving astrophysical black holes, the details of the expected X-ray emission spectra will depend both on the physics of the accretion disk and the properties of the black hole, including the theory of gravity from which they originate. Since black holes are not unique to General Relativity (GR), but rather they are a robust prediction of any metric theory of gravity, binary X-ray sources including black holes are excellent cosmic laboratories to test alternative theories of gravity assuming a model for the accretion disk.

\smallskip

Despite the fact that Einstein's General Relativity (GR) \cite{GR} has successfully passed many observational and experimental tests \cite{tests1,tests2,tests3}, alternative theories of gravity are well motivated for a number of reasons related to dark matter, dark energy, renormalizability etc. One can mention among others, for instance, $f(R)$ theories of gravity \cite{Sotiriou:2008rp,DeFelice:2010aj}, Ho\v rava gravity \cite{Horava:2009uw}, scale-dependent gravity \cite{SD1,SD2,SD3,SD4,SD5,SD6,SD7,SD8}, Weyl conformal gravity \cite{Kazanas:1988qa} and massive gravity \cite{deRham:2010kj}. 

In GR the static, spherically symmetric solution describing the gravitational field generated by a point mass is the Schwarzschild geometry \cite{SBH}, or the Schwarzschild-(anti) de Sitter space-time if a non-vanishing cosmological constant is allowed \cite{Bousso:2002fq}. However, additional terms are found in more complicated solutions for the metric tensor in gravity theories mentioned before. In the present work our goal is two-fold, namely we propose to study i) accretion process and ii) the X-ray spectrum expected from the disk in binary X-ray sources involving stellar-mass black holes, assuming a solution for the metric tensor obtained in theories other than GR, such as Weyl conformal gravity and massive gravity. In the first part or the article we shall discuss accretion of matter onto the black hole, while in the second part we shall show the expected X-ray emission spectra. In this particular solution for the metric tensor there is a additional term linear in the radial coordinate, and it is very appealing as it may explain galaxy rotation curves, see e.g. \cite{Mannheim:2010ti,Mannheim:2010xw,Panpanich:2018cxo}.

\smallskip

We organize the presentation of our manuscript as follows: After this introduction, in the next section 
we briefly present the key assumptions regarding the physics of the accretion disc. In section 3 we compute the most important quantities of interest, for the geometry assumed here. Finally, we finish with some concluding remarks in the fourth section.  We adopt the metric signature $(-,+,+,+)$, and we work most of the time in geometrized units where $k_B=c=\hbar=1=G$, using the conversion rules $1m = 5.068~10^{15}~GeV^{-1}$, $1kg = 5.61~10^{26}~GeV$, 
$1K = 8.62~10^{-14}~GeV$.

\section{Assumptions and physical considerations}

Here we shall formulate the problem, describing the assumptions and physical considerations that will allow us to compute the quantities of interest and present our main results in the next section.

\subsection{Accretion disk model}

The disk is characterized by a surface mass density $\Sigma(r)$, volume mass density $\rho(r)$ and height $2H(r)$. Then clearly
\begin{equation} \label{density}
\rho(r) = \frac{\Sigma(r)}{2H(r)}
\end{equation}

The properties of the disk depend on a number of factors, such as accretion rate, opacity, pressure etc.
For a nice review of the theory of black hole accretion disks see \cite{Abramowicz:2011xu}. In this work we adopt the standard model by Shakura-Sunyaev from the 70's \cite{Shakura:1972te}, suitable to describe geometrically thin, optically thick and cool accretion disks. Since the disk is cool the non-relativistic treatment is adequate here, for a fully relativistic treatment see \cite{PageThorne}. The main assumptions of the model adopted here, summarized very nicely in \cite{Faraji:2020skq}, are the following:

\begin{itemize}

\item[--] There are no magnetic fields.

\item[--]  Advection is negligible.

\item[--]  The accretion velocity $u^i$ has a radial component $u^r < 0$ only, 

\begin{equation}
u^\mu = (u^0, u^r, 0, 0).
\end{equation}

\item[--]  The disk is geometrically thin, i.e.

\begin{equation}
h(r) \equiv \frac{H(r)}{r} \ll 1
\end{equation}

\item[--]  The disk is optically thick: As the opacity, $\kappa$, is dominated by the Thomson scattering
\begin{equation}
\kappa=\sigma_T / m_p=0.4~cm^2/g
\end{equation}
where $m_p$ is the proton mass, and $\sigma_T$ is the Thomson cross section, the disc is opaque, characterized by a large optical depth, $\tau > 1$, and therefore it is optically thick.

\item[--]  The disk is cool: Quite generically, when matter is optically thick $\tau > 1$, the accretion disk can be quite luminous and also efficiently cooled by radiation. Radiation is important in accretion disks as a way to carry excess energy away from the system. In geometrically thin, optically thick (Shakura–Sunyaev) accretion disks, radiation is
highly efficient, and nearly all of the heat generated within the disk is radiated locally. Thus, the disk remains relatively cold, i.e.
\begin{equation}
T \ll \frac{M m_p}{r},
\end{equation}
where $T$ is the temperature of the disk, and $M$ is the mass of the black hole.

\item[--]  The equation-of-state (or total pressure), $P$, has two contributions, one from the gas and another from radiation

\begin{equation} \label{EoS}
P = P_{gas} + P_{rad} = \frac{T}{m_p} \: \rho + \frac{a}{3} T^4,
\end{equation}
with $a=4 \sigma$ being the radiation constant, and $\sigma$ being the Stefan-Boltzmann constant.

\item[--]  Accretion rate at the Eddingron limit: For a steady state disk, when a balance is reached between gravity, which acts inwards, and pressure, which acts outwards, the accretion rate remains a constant, $2 \pi r \Sigma u^r = -\dot{M}=constant$ given by the Eddington limit \cite{Faraji:2020skq} 
\begin{eqnarray}
\dot{M} & = & 16 L_{Edd} \\
L_{Edd} & = & 1.2 \times 10^{38} \left( \frac{M}{M_{\odot}} \right) erg/s
\end{eqnarray}

\end{itemize}

\begin{figure}[ht!]
	\centering
	\includegraphics[width=1.0\linewidth]{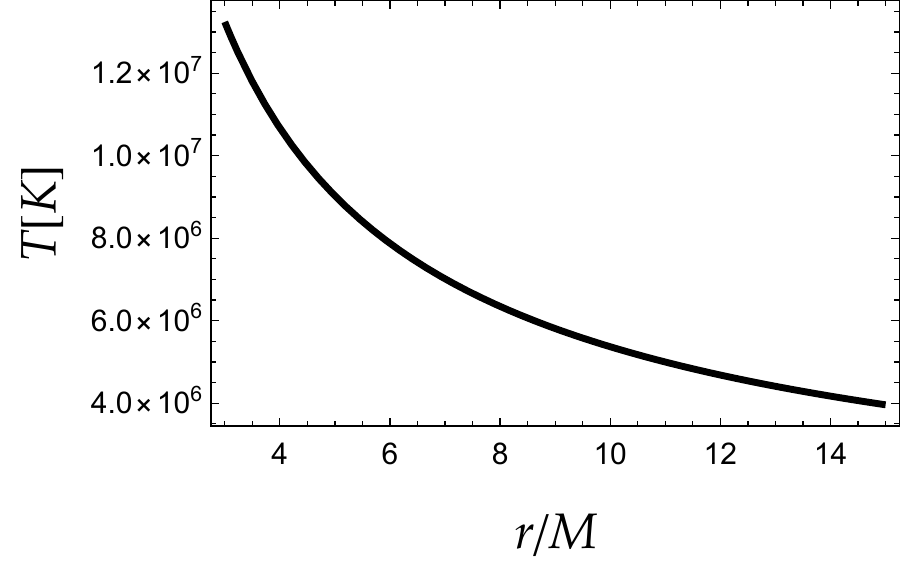}
	\caption{
		Temperature of the disc versus radial distance assuming $M=30~M_{\odot}$.
	}
	
	\label{fig:1} 	
\end{figure}

Those physically reasonable assumptions convert the problem into a system of algebraic equations easy to solve.
We see that once the temperature and the energy density are known, it is easy to compute the pressure. In the Shakura-Sunyaev model, the temperature obeys the following profile \cite{Shakura:1972te}
\begin{equation}
T(r) = \left[ \left( \frac{3M \dot{M}}{8 \pi \sigma r^3} \right) \left( 1-\sqrt{\frac{R_{in}}{r}}  \right) \right]^{1/4}
\end{equation}
where $R_{in}$ is the radius of the innermost stable circular orbit. Then, for distances $r \gg R_{in}$, the temperature takes a simpler power-law form
\begin{equation} \label{temperature}
T(r) = \left( \frac{3M \dot{M}}{8 \pi \sigma r^3} \right)^{1/4}
\end{equation}
i.e. it decays with the radial distance as $r^{-3/4}$, and it is easy to check that for an astrophysical black hole with a mass $M = (tens) M_{\odot}$, the temperature is of the order of $(10^6-10^7)~K$, see Figure \ref{fig:1} for 
$M=30~M_{\odot}$. As $T \ll m_p$, the assumption for a low temperature disk is clearly satisfied. 

Finally, we may compute the surface density and the semi-height, if we want to, making use of (\ref{density}) as well as the following equation \cite{Faraji:2020skq}
\begin{equation}
\Sigma = \frac{-\dot{M}}{2 \pi r u^r}
\end{equation}

The part related to the computation of the accretion velocity and the energy density is the subject of the next section.


\begin{figure*}[ht!]
	\centering
	\includegraphics[width=0.48\linewidth]{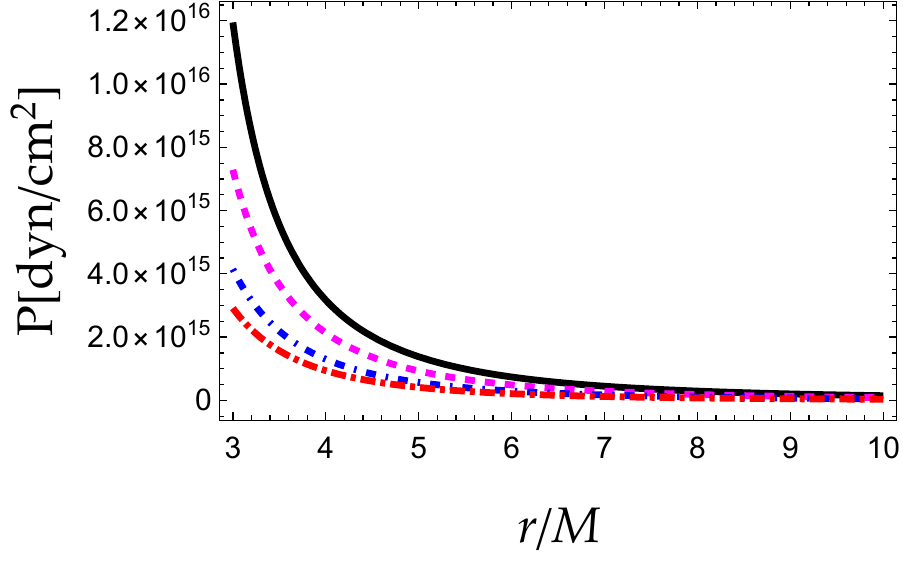} \ \
	\includegraphics[width=0.43\linewidth]{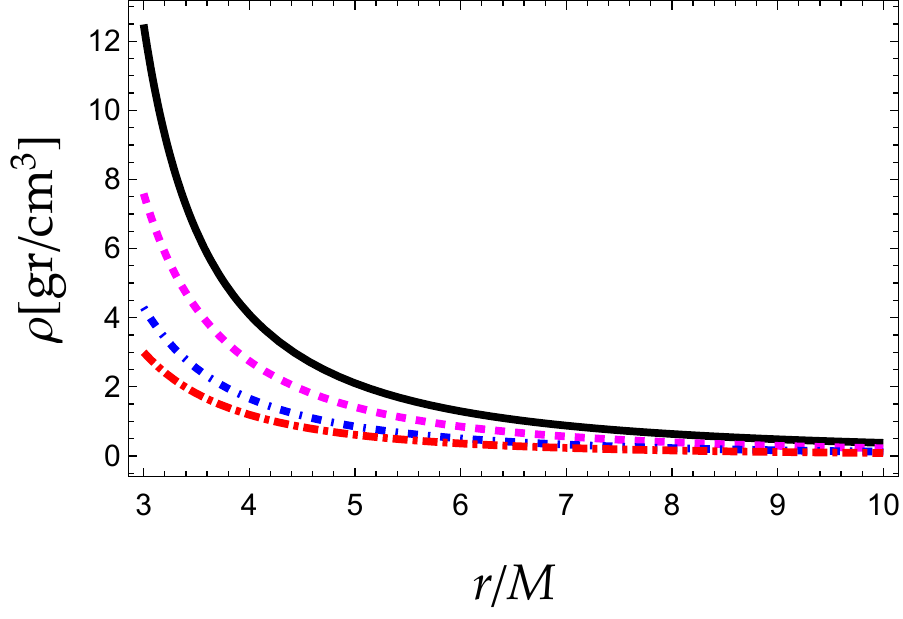} 
	\caption{
		Pressure (top) and mass density (bottom) versus radial coordinate assuming  $M=30~M_{\odot}$. 
		Shown are: $\epsilon=0$ (black),
		$\epsilon=0.05$ (magenta), $\epsilon=0.15$ (blue) and $\epsilon=0.25$ (red).
	}
	\label{fig:2} 	
\end{figure*}

\subsection{Accretion in spherically symmetric geometries}

A realistic black hole, since it is expected to be electrically neutral, it must be axisymmetric, Kerr-like \cite{kerr},
characterized by its mass, $M$, and rotation speed, $J$. In low-mass X-ray binaries, the spin parameter of black holes, $a^*=J/M^2$, covers the whole range, $0 \leq a^* \leq 1$, whereas in the few known cases of massive X-ray binaries, the spin of the black hole is found to be big \cite{kalogera,miller}. Therefore, in the following we shall consider static, spherically symmetric space-times in Schwarzschild-like coordinates $t,r,\theta,\phi$ of the form 
\begin{equation} \label{geometry}
ds^2 = -f(r) dt^2 + f(r)^{-1} dr^2 + r^2 d \Omega^2
\end{equation}
with some lapse function $f(r)$. 

Accretion processes for general spherically symmetric compact objects has been discussed in \cite{Bahamonde:2015uwa},
and after that several applications appeared in the literature, see e.g. \cite{paper1,paper2,paper3,paper4,paper5,paper6}. 
Recently relativistic dust accretion onto a scale-dependent polytropic black hole \cite{Contreras:2018dhs} was analyzed in \cite{Contreras:2018gct}. In the following we follow closely those two works \cite{Bahamonde:2015uwa,Contreras:2018gct}.

For matter we assume a perfect fluid with a stress-energy tensor of the form

\begin{equation}
T_{\mu \nu} = P g_{\mu \nu} + (\rho+P) u_\mu u_\nu
\end{equation}
where the velocity of the fluid satisfies the normalization condition 
\begin{equation}
g_{\mu \nu} u^\mu u^\nu = -1
\end{equation}
or, equivalently
\begin{equation}
-f (u^0)^2 + (1/f) u^2 = -1
\end{equation}

Using the two conservation equations, namely one for the energy and another for the number of particles \cite{Faraji:2020skq,Contreras:2018gct}
\begin{equation}
\nabla_\mu(\rho u^\mu)  =  0 , \;  \;  \;  \;  \;  \; \nabla_\nu T^{\mu \nu} = 0
\end{equation}
respectively, one obtains the following two equations involving $P, \rho$ and $u^0, u^r \equiv u$
\begin{eqnarray}
\rho u r^2 & = & C_1 \\
(P+\rho) u^0 u r^2 & = & C_2
\end{eqnarray}
where $C_1, C_2$ are constants of integration. When $P \ll \rho$ things are simplified, and we may easily combine the equations to obtain a single algebraic equation for the radial component, which is found to be
\begin{equation}
u(r) = - \sqrt{C_3 f(r)^2 - f(r)}
\end{equation}
where we have set $C_3 \equiv (C_2/C_1)^2$. In the final step, we go back to the original equations to compute 
the energy density
\begin{equation}
\rho(r) = \frac{C_1}{u(r) r^2}
\end{equation}

Therefore, the accretion velocity, $u^r=u$, and the energy density, $\rho$, are computed once the gravitational field, eq. (\ref{geometry}), is known. The precise form of the equation-of-state is not relevant at this step, as long as the pressure is negligible compared to the energy density, $P \ll \rho$. Finally, according to the physical considerations regarding the properties of the accretion disk, we can now compute the pressure using the above equation-of-state, eq. (\ref{EoS}). 
Clearly, there is an interplay between the theory of gravity via the lapse function in the last equations, and the physics of the disk, equations (\ref{EoS}) and (\ref{temperature}).

\section{Stellar mass black hole X-ray binary in Massive Gravity}

We are now ready to exploit the machinery developed in the previous section to actually compute
the main properties of a binary X-ray source involving astrophysical black holes in alternative theories of gravity.
In particular, we shall consider in the present work a space-time with a lapse function of the form
\begin{equation}
f(r) = 1 - \frac{2 M}{r} - \frac{\Lambda r^2}{3} + \gamma r 
\end{equation}
where $\Lambda$ is the cosmological constant, and $\gamma > 0$ is a new parameter. This is a solution to the field equations in contexts other than GR, such as Weyl conformal gravity \cite{Kazanas:1988qa} or massive gravity \cite{deRham:2010kj}. The cosmological constant term should correspond to the one that accelerates the Universe. Therefore, although in principle it is there, in the discussion to follow we shall neglect it since its observational value is extremely small. Thus, we consider a gravitational field of the form
\begin{equation}
f(r) = 1 - \frac{2 M}{r} + \left( \frac{\epsilon}{M} \right) \: r
\end{equation}
where we have ignored the cosmological term, and we have set $\gamma=\epsilon/M$, with $\epsilon$ being a dimensionless quantity.

We remark in passing that Weyl gravity a constant term, too, is necessarily there, whereas in massive gravity making an appropriate choice of the parameters of the model the constant term can be set to zero \cite{Panpanich:2018cxo}, which is precisely what we shall do here.

The event horizon is computed by $f(r_H)=0$, and it is found to be
\begin{equation}
r_H = \frac{-M + M \sqrt{1+8 \epsilon}}{2 \epsilon}
\end{equation}
recovering the usual horizon $r_0=2M$ corresponding to the Schwarzschild solution in the limit $\epsilon \rightarrow 0$.
In particular, when $\epsilon \ll 1$, the event horizon is given by the simple expression
\begin{equation}
r_H \approx r_0 (1-2 \epsilon), \; \; \; \; \; \; \epsilon \ll 1
\end{equation}
which demonstrates that the horizon of the new geometry is lower than the one corresponding to the Schwarzschild solution.

In Figure \ref{fig:2} we show both the pressure and the mass density as a function of the radial coordinate varying the parameter 
$\epsilon$ assuming an astrophysical black hole of mass $M=30~M_{\odot}$, and setting
\begin{eqnarray}
C_1 & = & 10^{-13} \\
C_2 & = & -(5/2) C_1
\end{eqnarray}

\begin{figure}[ht!]
	\centering
	\includegraphics[width=0.95\linewidth]{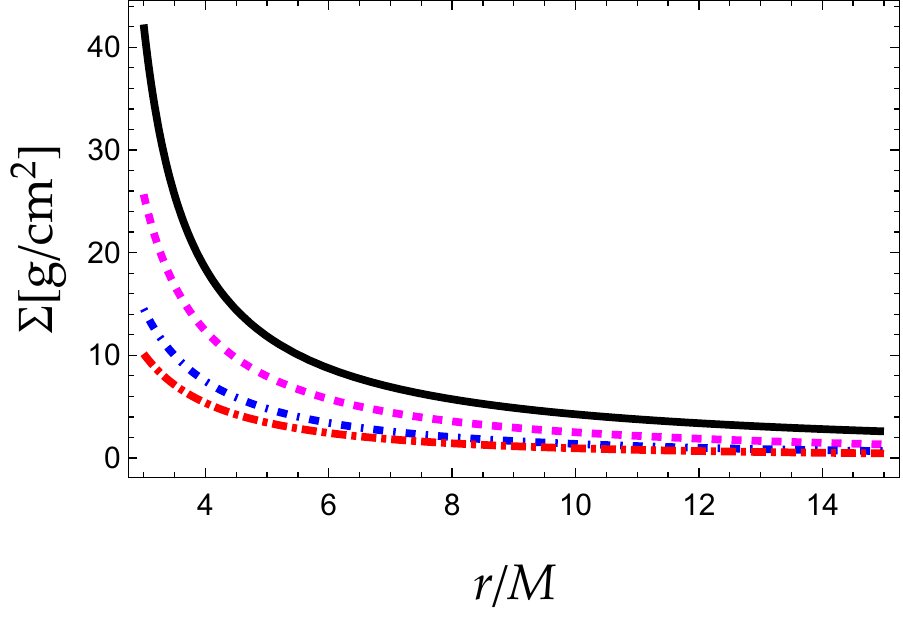} 
	\caption{
		Surface density versus radial coordinate, $\Sigma(r)$, assuming  $M=30~M_{\odot}$. 
		Shown are: $\epsilon=0$ (black),
		$\epsilon=0.05$ (magenta), $\epsilon=0.15$ (blue) and $\epsilon=0.25$ (red).
	}
	\label{fig:3} 	
\end{figure}

It is easy to verify that $h(r)=const.$ given by
\begin{equation}
h(r) = \frac{-\dot{M}}{4 \pi C_1} \sim 10^{-7} \ll 1
\end{equation}
and therefore one of the basic requirements is met. Finally, the semi-height and the surface mass density are given by
\begin{eqnarray}
H(r) & = & 10^{-7} r \\
\Sigma(r) & = & 2 \times 10^{-7} \rho(r) r
\end{eqnarray}
The semi-height grows linearly with $r$, while the surface density is shown in Figure \ref{fig:3}.


Figures 2 and 3 summarize our main results of the first part of our analysis regarding accretion of matter onto the black hole.

\subsection{Flux of X-ray emission}

Now we switch to normal units, and we quote the numerical values of the constants taken from particle data group:
$M_{\odot}  =  2 \times 10^{30}~kg $, 
$G =  6.67 \times 10^{-11}~m^3/(kg \: s^2)$, 
$c  =  3 \times 10^{8}~m/s $, 
$k_B  = 1.38 \times 10^{-23}~J/K $, 
$\sigma  =  5.67 \times 10^{-8}~W/(m^2 K^4)$, 
$h  =  6.63 \times 10^{-34}~J \: s$, 
$1~pc  =  3.09 \times 10^{16}~m $, 
$1~erg  =  10^{-7}~J$ and 
$1~eV  = 1.6 \times 10^{-19}~J $.

\smallskip

Regarding X-ray emission, the soft spectral component, $E \dot F_E$, expected from the optically thick disk, is given by \cite{Mitsuda:1984nv}
\begin{equation}
F_E = \frac{1}{D^2} \: cos(i) \: \int_{R_{in}}^\infty dr 2 \pi r B(E,T)
\end{equation}
where $B(E,T)$ is the Planckian distribution
\begin{equation}
B(E,T) = \frac{2 E^3}{c^2 h^3} \: \frac{1}{\text{exp}[E/(k_B T)]-1}
\end{equation}
where $h$ is the Planck constant, $c$ is the speed of light in vacuum, $k_B$ is the Boltzmann constant, $D$ is the distance from the source, and $i$ is the inclination ($0^o$ face-on, $90^o$ edge-on). Setting $s=r/R_{in}$ the flux is computed by
\begin{equation}
\begin{split}	
F_E = \left( \frac{R_{in}}{D} \right)^2 \: \cos(i) \\ \frac{4 \pi}{c^2 h^3} E^3 \int_1^\infty ds \frac{s}{\text{exp}[E/(k_B T(s))]-1}
\end{split}
\end{equation}
or, taking into account the spectral hardening factor, $f_{\small col}$, for a diluted blackbody spectrum \cite{Merloni:1999pe,Davis:2018hlj,Salvesen:2020bds}, the soft X-ray emission energy flux takes the final form
\begin{equation}
\begin{split}	
F_E = \left( \frac{R_{in}}{D f_{col}^2} \right)^2 \: \cos(i) \: \frac{4 \pi}{c^2 h^3} E^3
\\
 \int_1^\infty ds \frac{s}{\text{exp}[E/(f_{col} k_B T(s))]-1}
\end{split}
\end{equation}


\begin{figure}[ht!]
	\centering
	\includegraphics[width=1.0\linewidth]{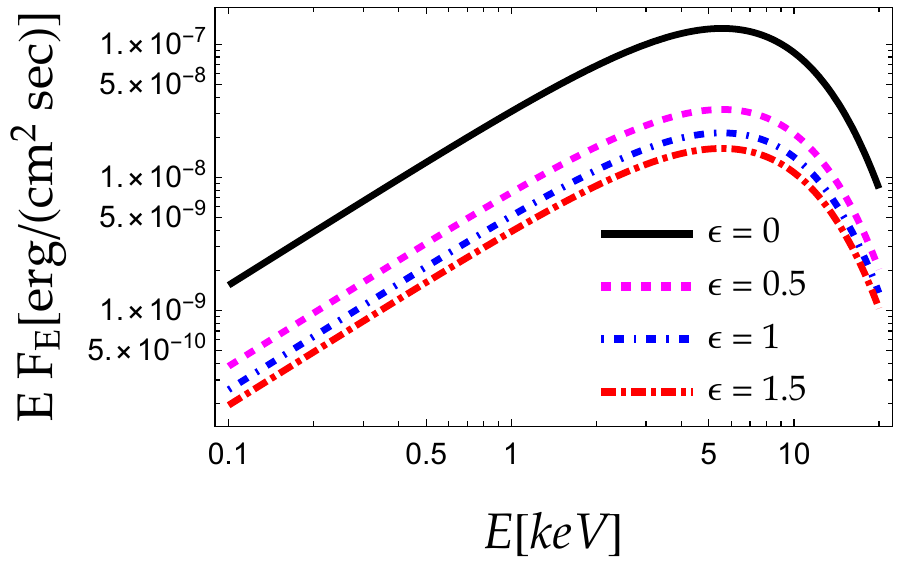}
	\caption{
		Soft X-ray emission from the black hole accretion disk assuming i) $D=10~kpc$, $i=70^o$, $f_{col}=1.6$ and 
		$M=10~M_{\odot}$, and $\epsilon=0, 0.5, 1, 1.5$ from top to bottom.
	}
	\label{fig:4} 	
\end{figure}


For a given configuration, i.e. known distance, inclination etc, the impact of the novel geometry on the expected spectrum is via the modification of the horizon. This is demonstrated in Figure \ref{fig:4} for a fictitious binary at distance $D=10~kpc$, inclination $i=70^o$, and $M=10~M_{\odot}$ to simulate some of the binaries shown in Table 1 of \cite{Salvesen:2020bds}. Fig.~4 summarizes our main result of the second part of our analysis regarding X-ray emission spectra 
expected from the disk. The spectral hardening factor varies over a narrow range, $f_{col} = (1.4-2)$ \cite{Davis:2018hlj}, while for the sources shown in Table 1 of \cite{Salvesen:2020bds}, it is found to be $f_{col} \approx 1.6$ in most of the cases. The computed spectrum as a function of the photon energy in a logarithmic plot exhibits the usual behavior with a peak at $E_* \sim (a few)~keV$. The additional linear term in the solution for the metric tensor within massive gravity shifts the curves downwards maintaining the position of the peak. Our results indicate that using observed spectra from confirmed X-ray binaries involving stellar mass black holes, we can put strong constraints on alternative theories of gravity.

\smallskip

We remark in passing that black hole solutions usually take advantage of certain oversimplifications, which are, among others, the connection between $g_{tt}$ and $g_{rr}$. In more complicated circumstances where the two metric potentials are independent, the normalization condition for the four-velocity of the fluid (used to obtain the final expression for the velocity profile) would alter the results obtained here. This is an interesting topic that motivates us to investigate the real impact of the assumption $g_{rr} \propto  g_{tt}^{-1}$ in alternative theories of gravity. This idea will be explored in another research paper.

\section{Conclusions}

In summary, in this article we have studied accretion disk and the soft spectral component of binary X-ray sources
in massive gravity. We have considered a binary system in which the primary object is an stellar mass black hole, while 
the companion (the donor) object is a sun-like star in the main or post-main sequence phase of evolution.
We model the optically thick, cool and geometrically thin accretion disk around the black hole adopting the physical considerations found in the seminal paper by Shakura and Sunyaev, according to which a few additional assumptions lead to a system of algebraic equations that are easily solved. In particular, in that model the temperature decreases with the radial distance as $r^{-3/4}$, while as far as the equation-of-state is concerned, the total pressure has two contributions, namely one from radiation and another from the gas. On the other hand, the gravitational field generated by the black hole is taken to be a static, spherically symmetric geometry (assuming a very low rotation speed) within Massive Gravity, where an additional term linear in the radial coordinate is found. That geometry is characterized by an event horizon which is find to be lower than the one corresponding to the Schwarzschild radius. First, using energy and number particle conservations, we computed the radial velocity and the energy density versus the radial coordinate. Then, given the temperature profile, the pressure was computed as well. Finally, regarding X-ray emission spectra, the soft spectral component expected from the disc is given by a blackbody-like spectrum, or to be more precise, it is represented as a superposition of multicolour blackbody spectra, since each layer of the accretion disc has a different temperature. For a given geometrical configuration (distance of the source, inclination etc) the spectrum depends on the event horizon of the geometry at hand. Thus, we have shown graphically the impact of the correction to the Schwarzschild geometry on the soft spectral component emitted from the disk. The computed spectrum as a function of the photon energy in a logarithmic plot exhibits the usual behavior with a peak at $E_* \sim 3 keV$. The additional linear term in the solution for the metric tensor within massive gravity shifts the spectrum downwards maintaining the position of the peak.

\section*{Acknowlegements}

The authors G.~P. and I.~L. thank the Funda\c c\~ao para a Ci\^encia e Tecnologia (FCT), Portugal, 
for the financial support to the Center for Astrophysics and Gravitation-CENTRA, Instituto Superior T\'ecnico, 
Universidade de Lisboa, through the Project No.~UIDB/00099/2020 and grant No. PTDC/FIS-AST/28920/2017.
The author A.~R. acknowledges Universidad de Tarapac\'a for financial support.


\end{document}